\documentclass[aps,twocolumn,nofootinbib,preprintnumbers,groupedaddress,letterpaper]{revtex4}

\usepackage[dvips]{color}
\usepackage[normalem]{ulem}
\usepackage{amsmath}
\usepackage{enumerate}
\usepackage{amsfonts}
\usepackage{epsfig}
\usepackage{yfonts}
\usepackage{undertilde}

\usepackage{graphicx}
\usepackage{bm}
\usepackage{subfigure}

\newcommand\comment[1]{}

\newcommand\poincare{Poincar\' e }

\newcommand\ov{\over }

\def\le{\left}
\def\ri{\right}

\def\({\left(}
\def\){\right)}
\def\[{\left[}
\def\]{\right]}
\def\<{\langle}
\def\>{\rangle}

\newcommand\half{{\ensuremath{\frac{1}{2}}}}
\newcommand\p{\ensuremath{\partial}}

\newcommand\field[1]{{\ensuremath{\mathbb{{#1}}}}}

\newcommand{\RR}{\field{R}}

\newcommand{\be}{\begin{equation}}
\newcommand{\ee}{\end{equation}}
\newcommand{\bea}{\begin{eqnarray}}
\newcommand{\eea}{\end{eqnarray}}
\newcommand{\bwt}{\begin{widetext}}
\newcommand{\ewt}{\end{widetext}}

\newcommand{\bi}{\begin{itemize}}
\newcommand{\ei}{\end{itemize}}
\newcommand{\ben}{\begin{enumerate}}
\newcommand{\een}{\end{enumerate}}
\newcommand{\bca}{\begin{cases}}
\newcommand{\eca}{\end{cases}}
\newcommand{\bln}{\begin{align}}
\newcommand{\eln}{\end{align}}
\newcommand{\bst}{\begin{split}}
\newcommand{\est}{\end{split}}

\begin{document}

\title{Pair-production of cusps on a string in AdS$_3$ }

\author{David Vegh} 
\email{d.vegh@qmul.ac.uk}
 
\affiliation{\it  Centre for Research in String Theory, School of Physics and Astronomy,
Queen Mary University of London, 327 Mile End Road, London E1 4NS, UK}


\begin{abstract}

The classical motion of a Nambu-Goto string in AdS$_3$ spacetime is governed by the generalized sinh-Gordon equation.
It can locally be reduced to the sinh-Gordon (shG), cosh-Gordon (chG), or Liouville equation, depending on the value of the scalar curvature of the induced metric. 
In this paper, I examine solutions that contain both shG-type and chG-type regions.
The boundary between these regions moves with the speed of light.
I show that near such boundaries (generalized) solitons  can be classically pair-produced. The solitons move subluminally (superluminally) in the shG (chG) region on the worldsheet, and they correspond to cusps on the string. A direct energy cascade is observed at the moment of pair-creation.

For the calculations, I use an exact discretization of the equation of motion. The solutions are segmented strings. In this discrete system, pair-production leads to a complete evaporation of the shG region. The final state is a gas of cusps in a chG environment. A Mathematica notebook that has been used to generate the relevant figures is attached.

\end{abstract}

\maketitle

\section{Introduction}

\begin{figure}[h]
\begin{center}
\includegraphics[width=7cm]{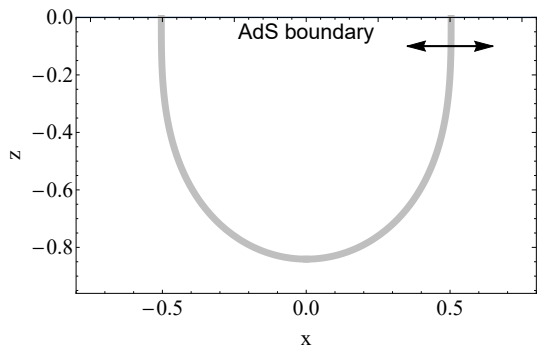}
\caption{\label{fig:setup} The setup: a string is hanging from the boundary of AdS$_3$. The configuration is initially static on the \poincare patch. Then, a propagating wave is created by moving one of the endpoints as shown by the arrows. After the quench, the endpoints are fixed again.
}
\end{center}
\end{figure}

This paper is concerned with the dynamics of a classical string in anti-de Sitter spacetime.  According to the gauge/gravity duality \cite{Maldacena:1997re, Gubser:1998bc, Witten:1998qj}, the string in the bulk is dual to a flux tube in the boundary field theory. In this paper, I will concentrate on the bulk side of the duality and study the string motion in AdS$_3$ spacetime.
Under certain conditions, the non-linear time-evolution creates pairs of cusps on the string.
The goal of the paper is to understand the qualitative features of this process in a concrete system: a string hanging from the boundary of AdS$_3$ spacetime. The initally static string configuration is shown in FIG. \ref{fig:setup}. In order to learn about the non-linear time-evolution, one of the endpoints is kicked, after which both endpoints are kept at a fixed position. This quench creates a propagating wave on the string.

For large enough quenches, direct energy cascades have been observed in the literature \cite{Ishii:2015wua}. Direct cascades transfer energy from large to small scales and thus ultaviolet physics becomes important. A prominent example is fully developed turbulence,  described by the Navier-Stokes equation. In that case, energy is dissipated at a scale set by the viscosity. In contrast, the string energy is conserved in AdS$_3$. In fact, since the motion is integrable \cite{Pohlmeyer:1975nb, DeVega:1992xc, Bena:2003wd}, there are infinitely many conserved quantities.
What happens in the UV instead of energy dissipation? As we will see, non-linear waves  ``break'' and pairs of cusps are produced on the string which then carry away the excess energy.
Such cusps (spikes) have been studied previously in \cite{Gubser:2002tv, Kruczenski:2004wg, Jevicki:2007aa, Jevicki:2008mm, Dorey:2008vp, Jevicki:2009uz, Dorey:2010iy, Dorey:2010id, Irrgang:2012xb} (see also the thesis \cite{Losi:2010hr} for a review).

The sigma model equations of motion can be converted into a generalized sinh-Gordon equation \cite{Pohlmeyer:1975nb}. By an appropriate change of coordinates, the equation locally reduces to the Liouville, the sinh-Gordon (shG), or the cosh-Gordon (chG) equation. This will be referred to as the {\it local type} of the equation. The type is selected by $\textrm{sign}(R-R_0)$, where $R$ is the scalar curvature of the induced metric, and $R_0$ is the constant negative curvature of elementary string segments.

Since the generalized shG equation encompasses both the shG and the chG equations, their relationship can be examined in a unified dynamical framework.
The main results of the paper are the following:
\begin{itemize}
\item  The cosh-Gordon theory has {\it superluminal} soliton solutions. These are the counterparts of the subluminal singular solitons of the sinh-Gordon theory.
\item  The sinh-Gordon theory can be thought of as a region in the cosh-Gordon theory where superluminal solitons have condensed. This region moves with the speed of light. 
\end{itemize}

Initially, the string is static (or time-periodic as we will see) and the local type is chG everywhere on the string. If the quench is sufficiently large, it can produce one or more shG-type regions. The boundary between the shG and chG regions travels with the speed of light.

One of the hallmarks of integrable field theories is the absence of particle production. How can cusps (which are solitonic objects at the level of the generalized shG equation) be pair-produced?
The answer lies in that they can only form near the boundaries between shG and chG regions.
One of the cusps falls into the shG region of the string, while the other cusp stays outside in the chG region. The cusp in the shG region is the well-known singular soliton of the shG equation. It should be thought of as a ``hole'' in the shG region. On the other hand, the chG theory has no known ``ordinary'' soliton solutions\footnote{The chG theory has no trivial solution which would be the starting point for generating multi-soliton configurations by  applying B\" acklund transformations.}. In this region, the cusp corresponds to a singular object with a spatial worldline: a superluminal soliton.


The string motion in AdS$_3$ is integrable. This allows for an exact discretization of the (1+1)-dimensional partial differential equation \cite{Vegh:2015ska, Callebaut:2015fsa, Vegh:2015yua, Gubser:2016wno, Gubser:2016zyw, Vegh:2016hwq, Vegh:2016fcm}. The discretization renders the equation of motion discrete in both space and time. The corresponding string solution is a segmented string. Any string solution can be approximated by segmented strings to arbitrary accuracy by increasing the number of segments and appropriately choosing the initial positions and velocities of the elementary segments.
Solving discrete equations has various advantages over numerical solutions of continuous PDEs. Most importantly, there are no numerical errors that would otherwise  accumulate over a long period of time.

We will investigate what happens to the string at late times after many oscillations.
If there is a shG region on the string, then cusps are constantly produced at its boundary.
Eventually, the shG region completely evaporates and leaves behind a gas of superluminal solitons in a chG environment. The evaporation time depends crucially on how many segments the string consists of (in the continuum limit the shG region does not shrink at all during the time-evolution).
The number of solitons in the final state and the original size of the shG region (measured in number of string segments) are equal. Thus, new particles have not been created, instead the shG region has disintegrated into a superluminal soliton gas.

The paper is organized as follows.
Section II outlines the derivation of the discrete equation of motion. Section III discusses how to set up the initial string. Even though this solution is static in the continuum limit, the segmented version is only time-periodic and requires some care. Section IV computes the non-linear string motion after the quench. The energy spectrum is calculated and an example for energy cascades is shown. Section V studies the cusp pair-production mechanism in detail. Section VI qualitatively describes the long-time dynamics: the evaporation of the shG region and the formation of the gas of cusps. Appendices A-C contain information about the technical details of the calculations. Appendix D describes an analytical string solution.

\section{The equation of motion}

\subsection{Classical string in AdS$_3$}

The canonical embedding of AdS$_3$ into $\RR^{2,2}$ is given by the universal covering space of the surface
\be
  \label{eq:surface}
  \vec Y \cdot \vec Y \equiv -Y_{-1}^2 - Y_0^2 + Y_1^2 + Y_2^2 = -1 .
\ee
Global AdS time is defined as the angle on the $Y_{-1}$, $Y_0$ plane. Figures in this paper will use coordinates\footnote{The coordinate transformation is given by
\be
  \nonumber
  (t, \, x, \, z) = \le(  {Y_{0} \over Y_{2} - Y_{-1}}, \ {Y_{1} \over Y_{2} - Y_{-1}}, \ {1 \over Y_{2} - Y_{-1}} \ri).
\ee} on the \poincare patch with metric
\be
  \label{eq:poinca}
  ds^2 = {-dt^2 + dx^2 + dz^2 \over z^2}
\ee
Here $z<0$ is the radial coordinate that will be shown as the vertical direction in the forthcoming figures. The reason for this choice is that in the $\{t,x,z\}$ coordinate system there is a gravitational force pulling massive objects down in the $z$ direction.
Hence, in order to study the long-term dynamics of the string, one has to hold it in place. The simplest way to do this is to keep both of its endpoints fixed, for instance, on the boundary. According to the AdS/CFT dictionary, the string is the holographic dual of the flux tube between two external quarks in the boundary field theory.

Our starting point is the Polyakov action 
\be
  \label{eq:action}
  S  = -{\tau_1 \ov 2}\int d^2 \sigma \, \sqrt{-h}h^{ab} \p_a X^M \p_b X^N  G_{MN} 
\ee
where $\tau_1$ is the string tension. $X^M$ are arbitrary coordinates on AdS$_3$, while $h$ and $G$ are the worldsheet and background metrics, respectively.

In terms of the $Y$ coordinates in (\ref{eq:surface}), the equations of motion in conformal gauge are
\be
\nonumber
  \p \bar\p \vec Y - (\p \vec Y \cdot \bar\p \vec Y ) \vec Y = 0 .
\ee
The second term comes from a Lagrange multiplier that keeps the string on the AdS$_3$ hyperboloid.
The equations are supplemented by the Virasoro constraints
\be
  \nonumber
  \p \vec Y \cdot \p \vec Y = \bar\p \vec Y \cdot \bar\p \vec Y = 0 .
\ee
and we have used the null worldsheet coordinates $z = \half(\sigma-\tau)$, $\bar z = \half(\sigma+\tau$).

\begin{figure}[h]
\begin{center}
\includegraphics[width=8cm]{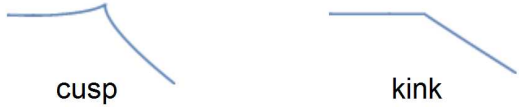}
\caption{\label{fig:cuspkink} Terminology used in the paper: cusps \& kinks on the string. At the location of cusps the string normal vector changes sign. They correspond to solitons in the sinh-Gordon theory and have a topological charge.
Kinks connect elementary string segments and move with the speed of light. There is no charge associated to these objects.
}
\end{center}
\end{figure}

The above system can be reduced to a generalized  sinh-Gordon theory \cite{Pohlmeyer:1975nb, DeVega:1992xc} by defining
\be
  \nonumber
  e^{2\alpha(z,\bar z)} = \half \p \vec Y \cdot \bar\p \vec Y
\ee
\be
  \nonumber
  N_a = \half e^{-2\alpha} \epsilon_{abcd} Y^b \p Y^c \bar\p Y^d
\ee
\be
  \nonumber
  p = -\half \vec N \cdot \p^2 \vec Y \ , \qquad  \bar p = \half \vec N \cdot \bar\p^2 \vec Y \ .
\ee
Note that $\vec N \cdot \vec Y =\vec N \cdot \p \vec Y = \vec N \cdot \bar\p \vec Y  = 0 $ and $\vec N \cdot \vec N = 1$. Furthermore, $p=p(z)$ and $\bar p = \bar p(\bar z)$ are functions of only one of the lightcone coordinates\footnote{In the Euclidean case, they are (anti-)holomorphic functions.}. The potential $\alpha$ satisfies the generalized sinh-Gordon equation
\be
  \label{eq:sinh}
  \p \bar\p \alpha(z,\bar z) - e^{2\alpha} + p(z)\bar p(\bar z) e^{-2\alpha} = 0 .
\ee
By the use of an appropriate coordinate transformation, the equation can be cast in the usual sinh-Gordon, cosh-Gordon, or Liouville equation form. The type of the canonical equation depends locally on the sign of $ p(z)\bar p(\bar z)$. Since $z$ and $\bar z$ are lightcone coordinates,  boundaries between shG and chG regions are also null. For instance, in the shG region
a new coordinate $w$ can be defined by
\be
  \nonumber
  dw = \sqrt{p(z)}dz 
\ee
In the $w$-plane the modified metric factor
\be
  \nonumber
 2 \hat \alpha(w,\bar w) = \alpha - {1\ov 4} \log p\bar p
\ee
satisfies the ordinary sinh-Gordon equation
\be
  \nonumber
  \p \bar\p  \hat\alpha - 4 \sinh \hat\alpha = 0 .
\ee

The shG teory has singular soliton and antisoliton solutions, given by the explicit formula
\be
  \label{eq:solitt}
{\alpha}_{s,\bar{s}}= \pm \log \le(\tanh^2  {\sigma-v\tau \ov \sqrt{1-v^2}}\ri)
\ee
\begin{figure}[h]
\begin{center}
\includegraphics[width=4cm]{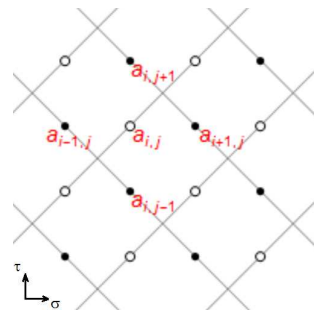}
\caption{\label{fig:sub} Kink worldlines form a quad lattice on the string worldsheet (parametrized by $\tau$ and $\sigma$). The field $a_{ij}$ lives on the edges (black or white dots depending on edge orientation).
}
\end{center}
\end{figure}

\noindent
where $v$ is the soliton velocity. Solitons with $\alpha \to -\infty$ correspond to {\it cusps} on the string\footnote{The motion of a Nambu-Goto string  on $\RR \times S^2$ can be described by the sine-Gordon equation \cite{Hofman:2006xt}. The sine-Gordon solitons are rather different from the singular solitons of the sinh-Gordon equation: the string embedding corresponding to a sine-Gordon soliton smoothly connects two points on the equator of $S^2$.}. On the other hand, antisolitons  with $\alpha \to +\infty$ correspond to places where the string touches the boundary of AdS$_3$. The boundary is infinitely far away which necessitates the blow-up of the metric factor.

At the location of cusps on the string, the normal vector $N \in \RR^{2,2}$ changes sign. Thus, cusps are topological objects which cannot be undone unless they collide with other cusps. As we will see in future sections, they appear in pairs when large non-linear waves on the string collapse.

Given a solution, the string embedding can be computed by solving an auxiliary Dirac equation where $\alpha$ appears as a potential.  Explicit formulas for solutions of the shG equation with $n$ solitons are known and the corresponding string embeddings have been found in \cite{Jevicki:2007aa}.
In this paper we will explore a different exact method which will be described in the following section.

\subsection{The discrete equation of motion}

A new route to solving the string equation of motion has recently been investigated \cite{Vegh:2015ska, Callebaut:2015fsa}. The classical theory is integrable which allows for the exact discretization of the equations. The solutions correspond to discrete string embeddings which consist of AdS$_2$ patches glued along null rays. (This is the generalization of a piecewise-linear string in flat space.) The string is smooth almost everywhere, but at the location of the {\it kinks} the normal vector jumps.

\begin{figure}[h]
\begin{center}
\includegraphics[width=5cm]{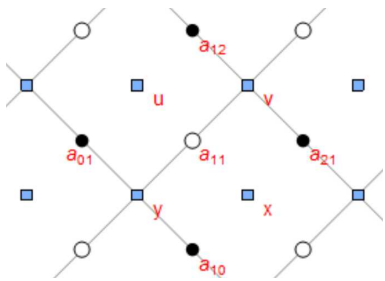}
\caption{\label{fig:fourpointeqn} Subdivided lattice. The discrete $a$ field has been extended to the vertices ($y,v$) and to the centers of plaquettes ($x, u$). }
\end{center}
\end{figure}

Kinks move with the speed of light on the worldsheet, and -- due to the Virasoro constraints -- in target space as well. This is necessary to maintain the ``segmentedness'' of the string.
The kink worldlines form a quad lattice on the worldsheet, see FIG. \ref{fig:sub}. Each diamond in the figure is a patch of AdS$_2$ with a constant normal vector. A discrete evolution equation for the normal vectors (or, equivalently, the kink collision points) has been found \cite{Vegh:2015ska, Callebaut:2015fsa} and can be used to build segmented string solutions.
The technique is ideally suited for long-time computations, because there are no accumulating numerical errors that typically scale with the lattice spacing.

In \cite{Vegh:2016hwq} I show that segmented strings in AdS$_3$ move according to the discrete equation of motion
\be
  \label{eq:deqn}
  \hskip -0.15cm {1\ov a_{ij} - a_{i,j+1}}+   {1\ov a_{ij} - a_{i,j-1}} =
  {1\ov a_{ij} - a_{i+1,j}}+   {1\ov a_{ij} - a_{i-1,j}}
\ee
Here $i$ and $j$ are integer indices labeling lattice points on the string worldsheet. As illustrated in FIG. \ref{fig:sub}, kink worldlines pass through each of the lattice points. These kinks move with the speed of light both in target space and on the worldsheet.
The points are colored alternatingly, depending on which way the corresponding kinks move.
The value of $a$ is related to the \poincare time where the extended null kink ray hits the AdS boundary.

The $a$ variables sitting on black and white dots correspond to advanced and retarded times, respectively. The discrete field $a$ is generically not constant along a kink worldline due to the collisions with other kinks.  

The kink lines surround a patch of AdS$_2$. The sum of all diamond-shaped patches is the entire segmented string worldsheet. It is an exact solution ({\it i.~e.} one does not need take the zero lattice spacing limit).

Segmented string embeddings can be obtained (up to a global $SL(2)$ transformation) from solutions of the equation (\ref{eq:deqn}). Details of the reconstruction procedure are described in Appendix A.

Finally we note that there exist other discretizations of the sinh-Gordon equation (see for instance \cite{Bastianello:2017zby} which uses an analytically continued version of Hirota's integrable discretization of the sine-Gordon model given by Orfanidis in \cite{PhysRevD.18.3822}).

\begin{figure}[h]
\begin{center}
\includegraphics[width=3cm]{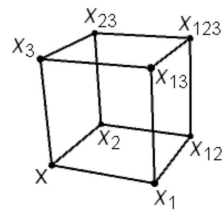}
\caption{\label{fig:3dcube} Higher-dimensional consistency. Using the 2d EOM on the sides of the cube, there are three ways to compute $x_{123}$ from $x, x_1, x_2, x_3$. If they give the same result, the 2d EOM is integrable.
}
\end{center}
\end{figure}

\subsection{Discrete integrability and 3d consistency}

All the calculations in the present paper have been done using eqn. (\ref{eq:deqn}). Although this equation has already a rather simple form, it can be cast into an even simpler one which will be explained in the following.  

Let us consider the square lattice in FIG. \ref{fig:sub} and subdivide it by adding extra variables to all the kink collision vertices and to the plaquette centers. The resulting lattice is displayed in FIG. \ref{fig:fourpointeqn}.
Let us now fix one of the new variables in the subdivided, say set $x=x_0$ in the figure. If we assume that the $a_{ij}$ field solves eqn. (\ref{eq:deqn}), then these values are given as well. We now demand that the cross-ratio of all small diamonds is equal to minus one. This will the new equation of motion for the subdivided lattice.
For instance, in the small diamond with the variables $\{ x, a_{10}, y, a_{11} \}$, we have
\be
  \label{eqn:crossratio}
  {(a_{10}-x)(a_{11}-y) \ov (x-a_{11})(y-a_{10})} = -1
\ee
From this equation, one can compute the value of $y$, since the other three are already given. By repeated use of this cross-ratio equation, one can fix all the other variables. It is easy to check that the cross-ratio equation is compatible with (\ref{eq:deqn}): going around $a_{ij}$ by applying the equation four times gives a consistent result.

For example, if we fix $x=x_0$, then first compute $y$, then $u$, and finally then $v$, then the cross-ratio equation with $\{x, a_{11}, v, a_{21} \}$ will be automatically satisfied.
The new variables in the lattice $x, y, \ldots$ will only depend on the value of $x_0$.

Integrability can easily be checked using the new form of the equation of motion (\ref{eqn:crossratio}): if the equation is such that the two-dimensional lattice can consistently be extended to a three-dimensional one, then the equation is integrable \cite{BobenkoSuris:2001}. This is shown in FIG. \ref{fig:3dcube}. Let us fix the values of $x, x_1, x_2$, and $x_3$. Using the cross-ratio equation, we can determine $x_{12}$, $x_{13}$, and $x_{23}$. Finally, there are three ways to compute $x_{123}$. If all of these give the same result, then the equation is integrable. It is easy to check that the cross-ratio equation indeed satisfies this criterion.

\clearpage

\begin{figure}[h]
\begin{center}
\includegraphics[width=7cm]{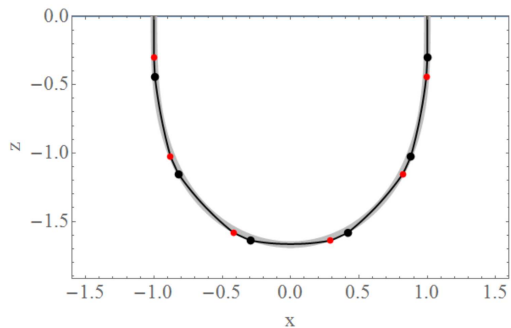}
\caption{\label{fig:static} Time-periodic segmented string with 12 kinks. It is an exact solution which approximates the static smooth solution in AdS$_3$ (background gray curve). Black and red dots indicate left- and right-moving kinks. These are connected by elementary segments (black lines).  
}
\end{center}
\end{figure}

\section{The initial setup}

Let us consider a static string in the bulk connecting two external quarks on the boundary of AdS$_3$. Due to time-translation symmetry, the string profile satisfies an ordinary differential equation which can be derived from the Nambu-Goto action. In terms of the \poincare patch coordinates (\ref{eq:poinca}) one gets
\be
  \nonumber
  2 x'(z)+2 x'(z)^3 - z x''(z) = 0
\ee
where $x(z)$ gives the location of the string in the $x$ direction as a function of the radial direction.

The solution to this equation can be expressed in terms of elliptic functions. The integration constants translate the solution in the $x$ direction, and rescale  the curve by an overall factor on the $x-z$ plane. The constants can be set such that the two string endpoints touch the boundary at $x=\pm 1$. The resulting string profile is plotted in FIG. \ref{fig:static} as a thick gray curve in the background.

We now need to set up a segmented string which approximates this static curve. The technical details of this are relegated to Appendix~B. At this point, the only parameter is the number of kinks (i.e. the number of elementary string segments minus one) which will be denoted by $N_\textrm{kink}$.
An example with $N_\textrm{kink}=12$ is displayed in FIG. \ref{fig:static}. Left- and right-moving kinks are shown in red and black dots, respectively. They are connected by black lines which indicate elementary string segments. The solution is not static: it oscillates slightly around the smooth solution in a time-periodic fashion as the kinks move and collide with the speed of light. The period is proportional to $1/N_\textrm{kink}$ (for large $N_\textrm{kink}$), since it depends on the average distance of neighboring kinks.

The smooth solution can be approximated to arbitrary precision by changing the number of segments (kinks). In the $N_\textrm{kink} \to \infty$ limit, the resulting string converges to the static solution.


\begin{figure}[h]
\begin{center}
\includegraphics[width=5cm]{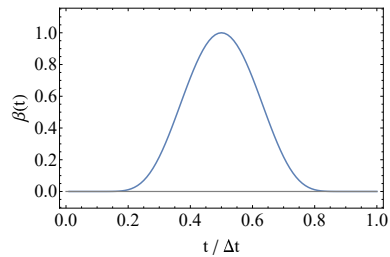}
\caption{\label{fig:quench} Compactly supported smooth function $\beta(t)$ for moving one of the string endpoints.
}
\end{center}
\end{figure}

\begin{figure}[h]
\begin{center}
\includegraphics[width=7cm]{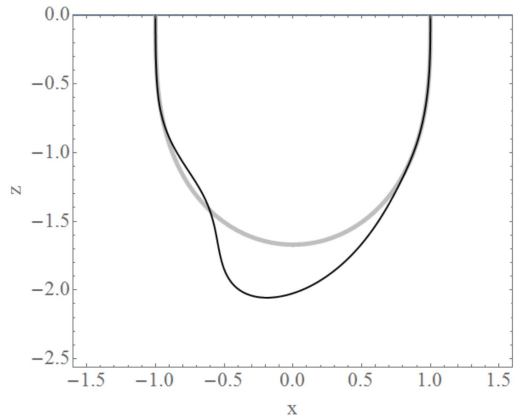}
\caption{\label{fig:perturbed} Segmented string with $N_\textrm{kink}=300$. The perturbation  of the left endpoint ($\Delta t=4$, $\epsilon=0.05$) has created a large wave on the string.
}
\end{center}
\end{figure}

\section{The Quench}

So far we have a time-periodic string embedding on the \poincare patch. The string endpoints $q$ and $\bar q$ are static quarks on the boundary.
In order to get something more interesting, let us move $\bar q$ so that the worldlines move according to the functions
\be
  \nonumber
  x_q(t) = -1 \qquad \textrm{and} \qquad x_{\bar q}(t) = 1+\epsilon \beta(t)
\ee
To make contact with earlier works in the literature, $\beta$ will be chosen to be the compactly supported $C^{\infty}$ function
\be
  \nonumber
 \beta(t)=
\begin{cases}
\exp\left[2\left(\frac{\Delta t}{t-\Delta t}-\frac{\Delta t}{t}+4
\right)\right]\qquad &(0<t<\Delta t)\\
0\qquad &(\textrm{otherwise})
\end{cases}
\ee
The function is plotted in FIG. \ref{fig:quench}.
The parameters of the quench are the amplitude $\epsilon$ and the length $\Delta t$. These must be chosen in a way so that the quark velocity never exceeds the speed of light.


\begin{figure}[h]
\begin{center}
\includegraphics[width=6cm]{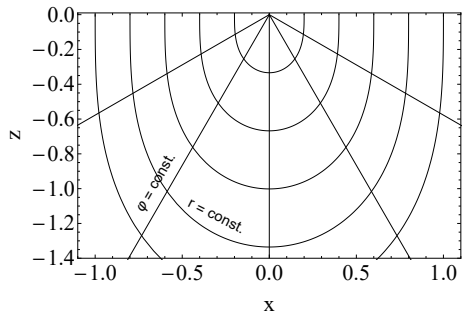}
\caption{\label{fig:polar} Polar-like coordinate system. Constant $r$ slices follow the shapes of static strings.
}
\end{center}
\end{figure}

The quench produces a large non-linear wave on the string and then moves the string endpoint back to its original position ($x_{\bar q} = 1$). Due to the Dirichlet boundary conditions, the wave bounces back and forth between the two endpoints. A snapshot with $\Delta t=4$ and $\epsilon=0.05$ is shown in FIG. \ref{fig:perturbed}. The initial static string embedding is a smooth gray curve, while the actual string configuration is drawn in black. In this example, the string consists of 300 segments. With such a large number of kinks, the embedding looks smooth. The number of segments is a constant: as a kink reaches a boundary, a new kink is sent in such that the Dirichlet boundary condition is maintained. The details of setting up the correct boundary conditions are given in Appendix~C.

The authors of \cite{Ishii:2015wua} kindly provided the numerical results of a simulation of the longitudinal one-sided quench. I have compared their numerical data to my exact results (using $N_\textrm{kink}=300$ and the same quenching function) and found a quantitative agreement between the two.


\section{Energy cascades} 

In this section we will compute the energy spectrum of waves on the string. Waves are defined by a displacement function which needs a reference embedding. Hence, the spectrum calculation is non-local and non-covariant. The reference embedding will be the initial, unperturbed static string which has been shown in gray in the figures.

The static string can be parametrized by a polar-like coordinate $(r, \varphi)$ which follows the shape of the static string. The new coordinate system is defined by \cite{Ishii:2015wua}
\be
  \nonumber
  z = r f(\varphi) \qquad x=  r g(\varphi)
\ee
where $f(\varphi)\equiv \textrm{sn}(\varphi; -1)$ is a Jacobi elliptic function and
\be
  \nonumber
   g(\varphi)\equiv -\int_{\beta_0/2}^{\varphi} d\varphi' f(\varphi')^2 \ ,
   \qquad \beta_0 \equiv {\Gamma({1\ov 4})^2 \ov 2 \sqrt{2\pi}}
\ee
In these coordinates, the static string embedding lies at $r(t, \varphi)= z_0 \equiv 2 \beta_0/\pi$ where $\varphi \in (0, \beta_0)$.

\begin{figure}[h]
\begin{center}
\includegraphics[width=7cm]{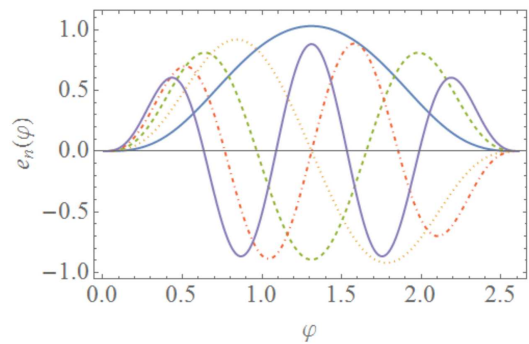}
\caption{\label{fig:modes} The first five eigenmodes of the linearized (continuous) string equation of motion as a function of the polar-like angle.
}
\end{center}
\end{figure}

The displacement function $\chi$ is defined by the perturbed embedding
\be
  \nonumber
   r(t, \varphi)= z_0 \[ 1+\chi(t,\varphi) \]
\ee
To linear order $\chi$ satisfies the equation \cite{Ishii:2015wua}
\be
  \nonumber
  \le(\p_t^2 - {1\ov z_0^2 h} \p_\varphi h \p_\varphi \ri) \chi = 0
\ee
where $h \equiv \le[ (g/f)' f  \ri]^2$.

Next, one can numerically determine the normal mode frequencies $\omega_n$ and eigenfunctions $e_n(\varphi)$ on the string. The first five modes are displayed in FIG. \ref{fig:modes}.
Any displacement function can be expressed in terms of these orthonormal eigenmodes as
\be
  \nonumber
  \chi(t, \varphi) = \sum_{n=1}^\infty c_n(t) e_n(\varphi)
\ee

This expansion is needed for computing the energy spectrum. Note that only single-valued functions can be expanded in this way. Once the string displacement is not such a function (for instance, there is a loop in the curve) the spectrum cannot be unambiguously computed.

\begin{figure}[h]
\begin{center}
\includegraphics[width=8cm]{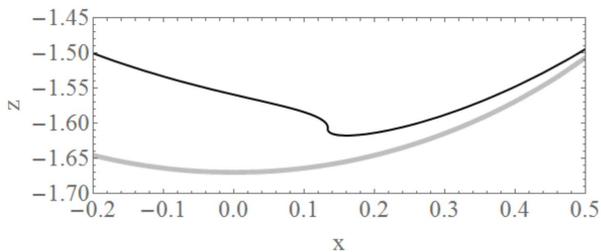}
\caption{\label{fig:breaking} Segmented string with 300 kinks after a quench ($\Delta t=8$, $\epsilon=0.095$). The wave bounces between the two endpoints and eventually it breaks.  At this point, a cusp/anticusp pair is created. The figure shows the moment when cusps are formed (at $t=22.59$).
}
\end{center}
\end{figure}

\begin{figure}[h]
\begin{center}
\includegraphics[width=8cm]{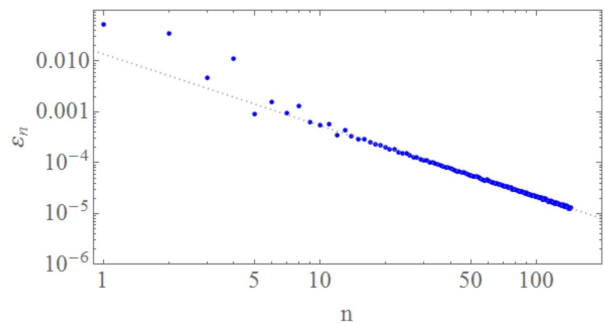}
\caption{\label{fig:cascade}  Energy spectrum right before the formation of the first cusp pair. A power-law fit gives $\varepsilon_n \propto n^{-1.41\pm 0.05}$. The parameters were $\Delta t=8$, $\epsilon=0.07$, $N_\textrm{kink}= 600$.
}
\end{center}
\end{figure}

Using the coefficients $c_n(t)$, the energy of a given mode is expressed as
\be
  \nonumber
  \varepsilon_n(t) \propto \le(  \dot c_n^2 + \omega_n^2 c_n^2 \ri)
\ee
In the linear theory $\varepsilon_n$ is conserved separately for each $n$. Even though the system is integrable and there are infinitely many conserved quantitites, the individual mode energies are not conserved in the full non-linear theory. The quench populates the low-lying modes while  higher modes are initially not filled. During the time-evolution energy is transferred from large to small scales, and the string embedding looks like a breaking wave as in FIG. \ref{fig:breaking}. In the figure, the static configuration is shown in gray while the actual string is a black curve. The steepest point is around $x \approx 0.13$ where the $\varphi$ derivative of the displacement function seems to diverge.
At this point in time, the spectrum obeys a power-law.  Such a power-law spectrum is presented in FIG. \ref{fig:cascade}.

A linear fit gives $\varepsilon_n \propto n^{-1.41\pm 0.05}$. The exponent does not seem to be universal. In fact, this is expected due to the covariance of the system: one cannot differentiate a priori between the background (the static string) and waves on top of it. By pumping in more energy, the string can be made arbitrarily long which changes the ``size of the box'' in which the waves propagate.

Shortly after the power-law spectrum is observed, the wave breaks  and two cusps form on the string.
Since  there is no viscosity in the system, there should exist another mechanism that gets rid of the excess energy at the ultraviolet end. This mechanism is the cusp pair-production process. Without cusp production, the energy must eventually flow back to the infrared. This seems to be the case when the string propagates in higher dimensions \cite{Ishii:2015wua}.

The breaking of the wave is somewhat similar to the way shockwaves form in solutions of the  inviscid Burgers equation (i.e. the (1+1)-dimensional Euler-equation without the pressure term). One important difference, however, is that the discrete string equation of motion (\ref{eq:deqn}) is not singular when the wave breaks and cusps form.

It is plausible that by turning on finite-$N_c$ effects (e.g. taking backreaction effects into account), viscosity and energy dissipation can be implemented. Cusps are likely to quickly lose energy by giving off a large gravitational radiation which can be detected at the AdS boundary. We will not investigate these possibilities here.

\section{Pair-production of cusps}

The Ricci scalar of the induced metric on the worldsheet can be expressed in terms of the metric factor.
\be
  \label{eq:ricci}
  R = -2 e^{-2\alpha} \p \bar\p \alpha
\ee
Let us rewrite the generalized shG-equation (\ref{eq:sinh}) as
\be
  \nonumber
  -2e^{-2\alpha} \p \bar\p \alpha(z,\bar z)  - 2 p(z)\bar p(\bar z) e^{-4\alpha} = -2 .
\ee
In the first term we recognize the scalar curvature $R$. Since $ e^{-4\alpha}>0$, we have
\be
  \nonumber
   \textrm{sgn}(R-R_0) = \textrm{sgn} \, p\bar p
\ee

\begin{figure}[h]
\begin{center}
\includegraphics[width=2.2cm]{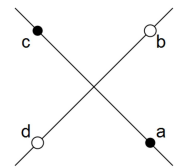}
\caption{\label{fig:deficit} The collision of two kinks on the segmented string worldsheet generates a deficit angle. The sign of the angle selects the local type of the equation of motion (shG or chG).
}
\end{center}
\end{figure}

\noindent
where $R_0 \equiv -2$ is the constant curvature of elementary string segments. Hence, by computing the scalar curvature, one is able to decide whether (\ref{eq:sinh}) can be reduced to the shG or to the chG equation via an appropriate coordinate transformation. This is important, because the two equations behave differently. The shG-equation has singular soliton solutions that can be generated by the B\" acklund transformation. This procedure is not available for the chG-equation, since its potential is unbounded from below and one cannot start adding solitons to its ground state.
The difference in the potentials also has consequences for the string dynamics. For instance, in the shG case, the cosh potential blows up as $\alpha \to -\infty$ which stops strings from collapsing.

On the segmented string worldsheet $R= R_0$ almost everywhere, except for those points where kinks collide. The local behavior of the equation (shG or chG) is  dictated by the sign of the integrated Ricci scalar  that is produced by the collision. It is equal to twice the deficit angle and is given by the formula \cite{Vegh:2016fcm}
\be
  \nonumber
  2\phi_\textrm{vertex}  =
  \int_\textrm{vertex} d^2 \sigma \sqrt{-g} \, R = 2 \log \le[ {(b-c)(a-d) \ov (c-d)(b-a)} \ri]^2
\ee
where $a,b,c,d$ are the discrete field variables around the vertex (see FIG. \ref{fig:deficit}).

\bwt

\begin{figure}[h]
\begin{center}
\includegraphics[width=18cm]{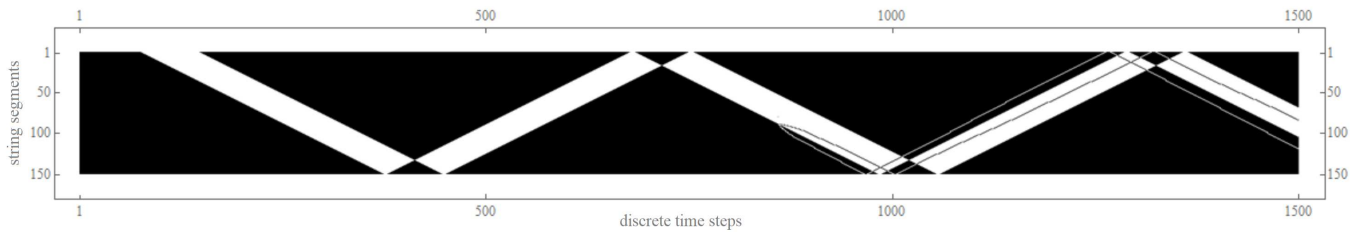}
\caption{\label{fig:ricci1} Local type of the generalized sinh-Gordon equation on the worldsheet. White (black) color indicates that the equation can be reduced to the ordinary sinh-Gordon (cosh-Gordon) equation. The parameters of the simulation are $N_\textrm{kink}= 300$, $\Delta t=8$, and $\epsilon=0.095$. After 860 discrete time  steps, a cusp-anticusp pair emerges.
}
\end{center}
\end{figure}

\begin{figure}[h]
\begin{center}
\includegraphics[width=18cm]{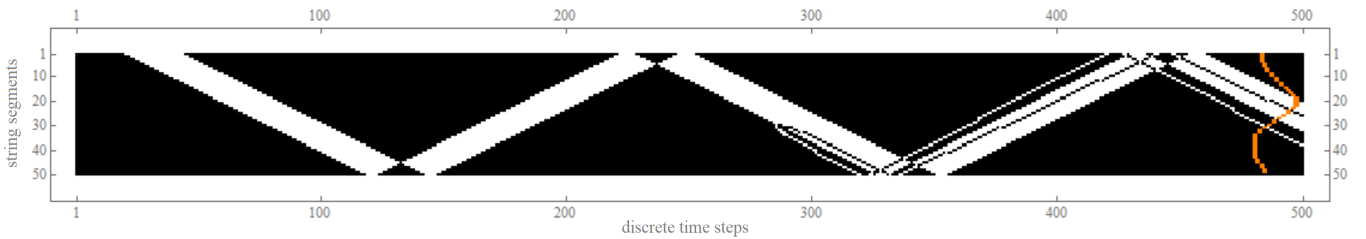}
\caption{\label{fig:ricci2} The parameters of the simulation are $N_\textrm{kink}= 100$, $\Delta t=8$, and $\epsilon=0.095$. The orange curve is a \poincare timeslice. The corresponding string embedding is shown in FIG.~\ref{fig:cusppair}.
}
\end{center}
\end{figure}

\ewt

The sign of the deficit angle can be plotted as a function of the discrete worldsheet variables. These plots are seen in FIG. \ref{fig:ricci1} and \ref{fig:ricci2}. Black (white) color corresponds to chG (shG) regions on the worldsheet. The horizontal direction is worldsheet time $\tau$ and the vertical axis is the spatial $\sigma$ coordinate. The only difference between the two figures is the number of string segments used in the simulation ($N_\textrm{kink} = 300$ and 100, respectively). Although the second figure is more pixellated, there is no significant difference. The quench creates a (white) shG region that bounces off the string endpoints (the bottom and top of the plot).

After several discrete time steps, a cusp pair forms near the leading boundary of the shG region. This happens at after 860 steps in the $N_\textrm{kink} = 300$ case, and after 290 steps in the $N_\textrm{kink} = 100$ case. The corresponding \poincare times are almost the same in the two cases. This shows that the segmentedness of the string is not relevant for cusp-production.

\begin{figure}[h]
\begin{center}
\includegraphics[width=7cm]{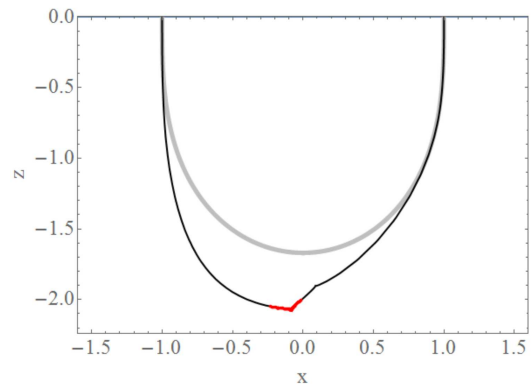}
\caption{\label{fig:cusppair} Cusp and anticusp on the string. The shG and chG regions are indicated by red and black, respectively. The anticusp lies in the shG region (at $x \approx -0.1$) and the cusp is located in the chG region (at $x \approx 0.1$). The initial static configuration before the quench is shown in gray.
}
\end{center}
\end{figure}

The cusp pair is shown in FIG. \ref{fig:cusppair}. Note that in the shG region the cusp corresponds to a subluminal singular soliton. In the chG region the worldline of the cusp on the worldsheet is superluminal. The corresponding singular solutions of the chG theory will be dubbed superluminal solitons.

The reason for superluminality can be understood by zooming in on the solitons. In such a limit, the generalized shG-equation reduces to a Liouville-like form
\be
  \nonumber
  \p \bar\p \alpha + p\bar p e^{-2\alpha} = 0 .
\ee

\noindent
A sign change of $p\bar p$ can be compensated by swapping the worldsheet coordinates $\tau \leftrightarrow \sigma$. This changes the sign of the first term and renders subluminal soliton worldlines superluminal.

Although the soliton is superluminal in the chG region, the corresponding cusp is perfectly causal in target space. FIG. \ref{fig:cusppair}  shows that the cusp in  the chG region  is pointing inward, therefore its tip can travel slower in target space, because it's moving on a curve of smaller ``radius''.  

\bwt

\begin{figure}[h]
\begin{center}
\includegraphics[width=14cm]{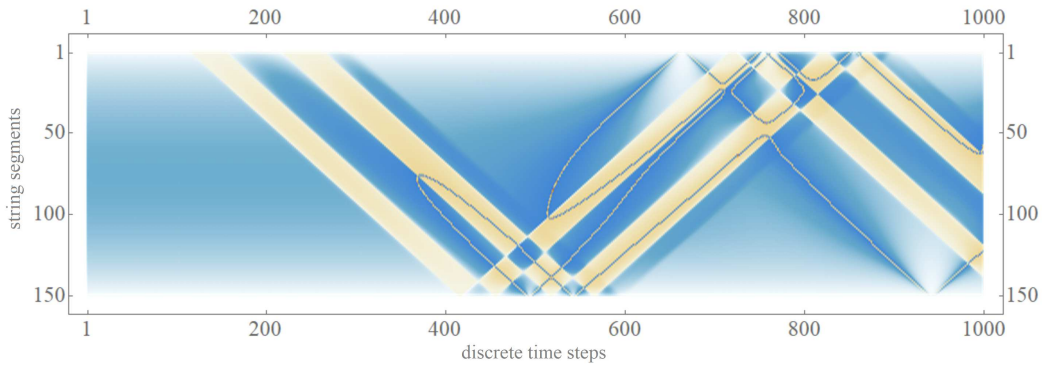}
\caption{\label{fig:colorricci}  Density plot of the deficit angle on the segmented string worldsheet. The quench creates two sinh-Gordon regions (shown in yellow). The time-evolution produces several cusp pairs, some of which annihilate.
The parameters of the simulation are $\Delta t=4$, $\epsilon=0.095$, with $N_\textrm{kink} = 300$.}
\end{center}
\end{figure}

\ewt

The results of another simulation are shown in FIG. \ref{fig:colorricci}. The amplitude of the quench remains the same, but the quench time is half as long, therefore the endpoint moves twice as fast. This motion creates two shG regions on the worldsheet. The figure is a density plot of the induced curvature. Superluminal soliton worldlines are surrounded by blue ``curvature clouds'' whose size depends on the (superluminal) speed of the soliton. As the soliton speeds approach the speed of light near the string endpoints, the cloud size vanishes.

\begin{figure}[h]
\begin{center}
\includegraphics[width=6cm]{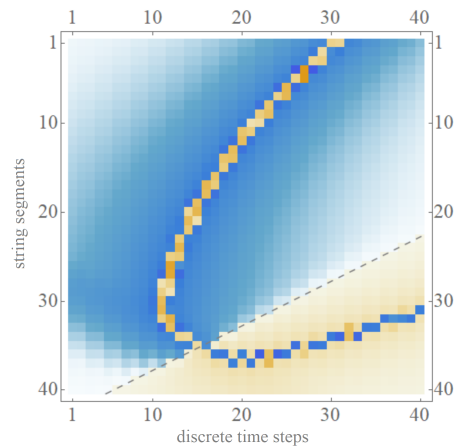}
\caption{\label{fig:cuspcreation} Density plot of the deficit angle on the segmented string worldsheet. The shG (chG) region is shown in yellow (blue). Dashed line indicates the null boundary between the two regions.
A cusp pair is produced near the boundary. The parameters of the simulation are $\Delta t=4$, $\epsilon=0.1$, $N_\textrm{kink} = 300$.
}
\end{center}
\end{figure}

FIG. \ref{fig:cuspcreation} is a close-up plot highlighting the region of cusp production on the worldsheet. One of the cusps falls into the shG region, while the other one stays outside in the chG region. Cusps are always produced near the leading front of the shG interval. The worldline of the singularity crosses the boundary (dashed line) with the speed of light. This is necessary, since on one side it is superluminal, while on the other side it has to move subluminally.

Note that $\hat \alpha$ blows up at the boundary of the region where $p\bar p \to 0$.  

Solitons that fall in slowly sink to the rear end of the shG region where they either annihilate with another incoming soliton, or just delete one segment from the shG region.  In the target space, the corresponding cusps completely disappear (this is related to the fact that at the boundary of the shG region we have $p\bar p \to 0$ and the equation degenerates into the Liouville equation).

The authors of \cite{Ishii:2015wua} have noted that the cusp formation time becomes longer as the quench amplitude $\epsilon$ gets smaller. They have performed a numerical extrapolation and have found a threshold value below which no cusps form. In our language, the reason for this is rather simple: for small enough values of $\epsilon$ the quench does not create a sinh-Gordon region on the string. Such a region is necessary for cusp formation. I have computed the threshold values of $\epsilon$ (at which the shG region consists of a single string segment) for a few values of $\Delta t$, and the results matched the numerical results in \cite{Ishii:2015wua}.

Note the ``fermionic'' nature of cusps on the string. At the tip the normal vector $N$ changes sign. Thus, one cannot place two cusps on top of each other, since flipping the sign twice corresponds to nothing at all.
Cusps repel each other which is visible at the level of the shG equation: soliton worldlines never cross.

We end this section by referring the reader to Appendix D which describes an analytical string embedding in (2+1)-dimensional flat spacetime. Even though the solution is periodic in time, it shows some similarities to the AdS$_3$ solutions discussed previously.
In particular, analogs of shG and chG regions can be identified. In flat space these are selected by $\textrm{sgn} \, R$.

\clearpage

\bwt

\begin{figure}[h]
\begin{center}
\includegraphics[width=18.5cm]{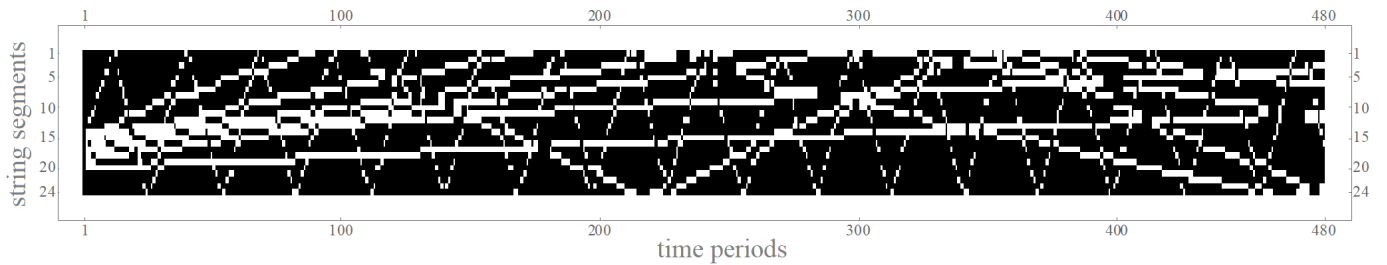}
\caption{\label{fig:evap0} Recurrence plot of cusps on the worldsheet.
The parameters of the simulation are $\Delta t=8$, $\epsilon=0.1$, $N_\textrm{kink} = 50$. Each pixel column in the plot is separated by the oscillation period (i.e. $2N_\textrm{kink}$ discrete time steps) under which the  sinh-Gordon interval (shown in white) bounces off both string endpoints and moves back to the same position. The interval quickly disintegrates after a few periods.  
}
\end{center}
\end{figure}

\begin{figure}[h]
\begin{center}
\includegraphics[width=18cm]{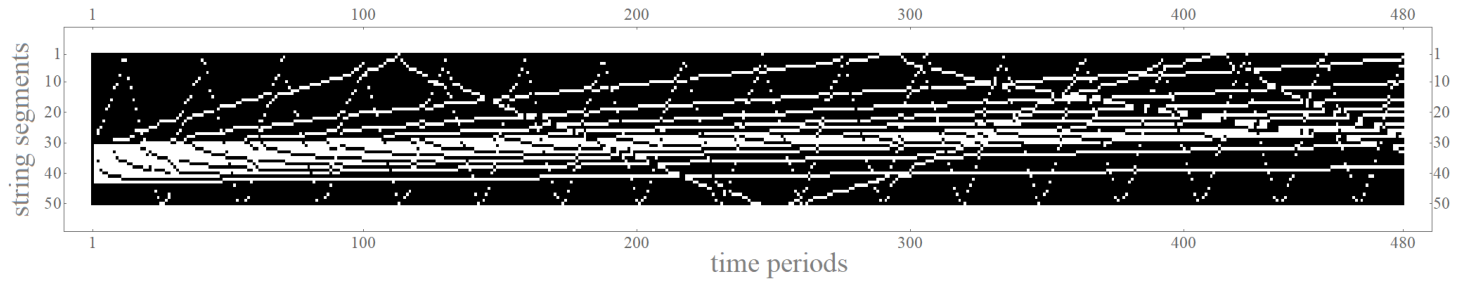}
\caption{\label{fig:evap1} Recurrence plot of cusps on the worldsheet. The parameters of the simulation are $\Delta t=8$, $\epsilon=0.1$,  $N_\textrm{kink} = 100$.
}
\end{center}
\end{figure}

\begin{figure}[h]
\begin{center}
\includegraphics[width=14.0cm]{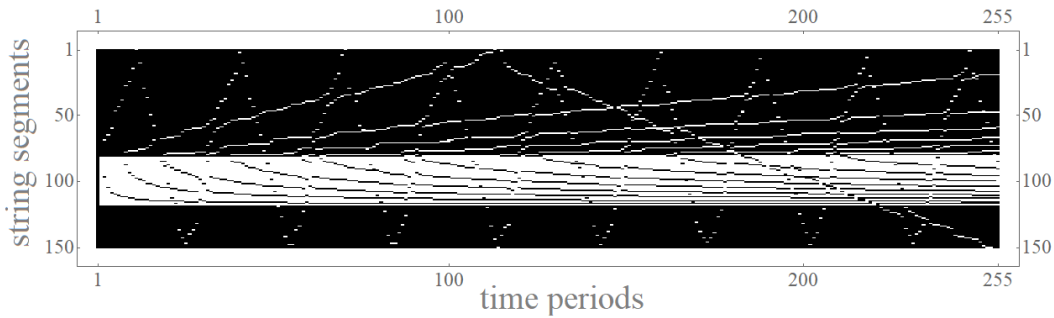}
\caption{\label{fig:evap2} Recurrence plot of cusps on the worldsheet. The parameters of the simulation are $\Delta t=8$, $\epsilon=0.1$,  $N_\textrm{kink} = 300$.
}
\end{center}
\end{figure}

\ewt

\section{Long-time dynamics}

The production of cusps eventually leads to the complete evaporation of the shG region at least in the segmented case. This is highlighted in FIGs. \ref{fig:evap0}-\ref{fig:evap2}. The three figures show the local type of the generalized shG-equation. The type is computed from the sign of the deficit angle at the kink collision vertices. Black and white colors indicate chG and shG regions, respectively. The vertical axis labels string segments, it is the discrete analog of the continuous worldsheet coordinate $\sigma$. The horizontal coordinate does not correspond to discrete time steps as in previous figures. In order to be able to display large time scales, we wait a full oscillation period between two adjacent columns. During this time, the wave bounces off both string endpoints and moves back to its original position. Correspondingly, the white shG interval also moves back to the original position on the string.
The time period (expressed in discrete time steps) is twice the total number of string segments since the shG interval propagates with the speed of light.

\clearpage

\begin{figure}[h]
\begin{center}
\hskip -0.2cm \includegraphics[width=9cm]{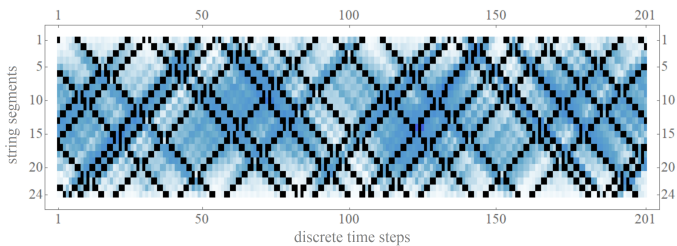}
\caption{\label{fig:gas} Density plot of the deficit angle on the worldsheet. The horizontal axis labels elementary time steps.  The parameters of the simulation are $\Delta t=8$, $\epsilon=0.1$, $N_\textrm{kink} = 50$, after 50000 time steps (not shown). The figure corresponds to the last two columns in FIG. \ref{fig:evap0}.
}
\end{center}
\end{figure}

The recurrence plots give us an idea about the behavior of the system over long simulation times. FIG. \ref{fig:evap0} indicates that the shG region completely evaporates after approx. 100 periods. After the evaporation, the string only contains a chG region (similarly to the static string before the quench) plus a gas of superluminal solitons.
The cusps that fell into the shG interval are gone: they behave like holes in that region.

In FIG. \ref{fig:evap1}, the string has twice as many elementary segments and thus the effective ultraviolet cutoff is also twice as large. Hence, the shG region stays together for a longer period of time.

The behavior of cusp pairs is best seen in FIG. \ref{fig:evap2}. This case has the largest ``resolution'' with  $N_\textrm{kink} = 300$ which means that the simulation approximates a smooth string to a greater accuracy.
A cusp that is thrown out into the chG region oscillates between the two string endpoints. The other cusp that fell into the shG region slowly sinks to the bottom end of the interval where it changes the type of the last segment into chG. At this point the cusp completely disappears. Presumably this has to do with the fact that at the boundary the generalized shG equation degenerates into the Liouville equation and the cusp size vanishes in this limit.

Note that as time passes, the worldlines of newly minted cusps are increasingly closer to being lightlike. Furthermore, it seems that the rate at which cusps are produced is decreasing over time. A quantitative understanding of the phenomenon would be needed to express the evaporation time as a function of $N_\textrm{kink}$.

Based on the figures, it is natural to extrapolate and speculate about the continuum limit of smooth strings. It seems that in that case the shG region never evaporates, since the segment size is zero, but the cusp production rate is roughly independent of the number of kinks.

FIG. \ref{fig:gas} displays a density plot of the scalar curvature in the final state (a sample of 200 time steps after the evaporation). Black lines indicate superluminal solitons of the chG theory where the curvature blows up.

Finally, I note that on even longer time scales the solitons can condense again forming a smooth shG region. The study of such recurrences was beyond the focus of the paper.

\section{Discussion}

This paper has studied the non-linear motion of a string in AdS$_3$ spacetime. The motion is governed by an underlying generalized sinh-Gordon equation which can be reduced to the sinh-Gordon, cosh-Gordon, or Liouville equation. The dynamics of small strings changes in the various cases, since the associated potentials are different. Sinh-Gordon strings for instance always reach a minimal size because the cosh potential blows up for small values of the field. Cosh-Gordon strings on the other hand can collapse \cite{Larsen:1996gn}.

The sinh-Gordon equation has well-known singular soliton solutions which correspond to cusps on the string. In this paper we have discovered analogous (albeit superluminal) soliton solutions of the cosh-Gordon equation.

The calculations were performed using an exact discretization of the equations. The solutions are segmented strings, built from elementary pieces. In a flat target space limit these pieces are straight lines. Instead of using the reflection formulas of \cite{Vegh:2015ska, Callebaut:2015fsa}, the computation relied on the variables introduced in \cite{Vegh:2016hwq} which were better behaved numerically. Although the calculations are exact in both cases, repeated application of the reflection formula exponentially magnifies the tiny numerical errors that arise from the fact that computers store real numbers only up to a certain number of digits.

By setting up the initial and boundary conditions carefully, the string worldsheet can contain regions of different local equation type.
(Note that for a segmented string, the local type is almost everywhere Liouville. What one means by sinh-Gordon or cosh-Gordon type is measured at points on the worldsheet where two kinks collide. The collision generates a deficit angle on the worldsheet and the sign of the angle gives the local type: shG or chG).

We have seen that the sinh-Gordon region evaporates via the cusp pair-production mechanism. A pair of cusps is produced on the string just outside the null boundary between the shG and chG regions. One of them falls into the shG region where it corresponds to the well-known singular soliton solutions of the shG theory. The other one stays outside in the chG region. Its worldline is superluminal and the corresponding object in the cosh-Gordon theory is a superluminal soliton.

There are many ways to extend this work:  

\begin{itemize}

\item
Long strings play an important role in the AdS/CFT correspondence.
It would be interesting to understand segmented strings from the boundary point of view. Closed segmented strings in the bulk should correspond to a distinguished class of operators of large canonical dimension in the CFT.

\item
Although we have shown the existence of superluminal solitons of the cosh-Gordon theory, we have found no analytical expressions describing these solutions which are valid in a smooth limit. It may be possible to obtain such formulas if the background is carefully chosen.

\item
The generalized sinh-Gordon equation is relevant for the 2d Gross-Neveu model since it governs the evolution of the mesonic mean field \cite{Neveu:1977cr}.
Boundedness of the meson field selects asymptotically shG-type solutions. Previous work has focused on purely shG-type solutions (see e.g. \cite{Klotzek:2010gp}). Generically, however, they can contain intervals of chG regions as well. It would be interesting to understand the consequences of our results in this context.

\item
Cusps are topological objects only in three spacetime dimensions. In four dimensions, for instance, they only appear for a moment on the string. In AdS$_n$, the string motion is integrable and using a higher-dimensional version of the spinor-helicity formalism, it is possible to obtain discrete equations of motion in various dimensions.

\item
This paper was only concerned with open strings. It is plausible that similar phenomena also exist for closed strings. Since these compact objects eventually fall through the \poincare horizon, global AdS coordinates are better suited for their study.

\item
It is possible to couple the string to a certain background B-field while preserving integrability of the equations \cite{Vegh:2016fcm}. For a special value of the coupling one arrives at the WZW model. We expect similar behavior in these systems, especially since the underlying discrete equations are the same.

\item
It would be interesting to study the thermodynamics of the system.

\end{itemize}

I hope to address some of these points in the future.

\vspace{0.2in}   \centerline{\bf{Acknowledgments}} \vspace{0.2in}
I am supported by the STFC Ernest Rutherford grant ST/P004334/1.
This work was initiated at the Center of Mathematical Sciences and Applications at Harvard University. The calculations were performed while I was a Delta-ITP fellow at Utrecht University.
I thank Garrett Goon, Diego Hofman, Takaaki Ishii, Martin Kruczenski, Wilke van der Schee for valuable discussions, correspondence, and comments on the manuscript.


\appendix
\section*{Appendix A: Reconstructing the embedding}

\begin{figure}[h]
\begin{center}
\includegraphics[width=3.5cm]{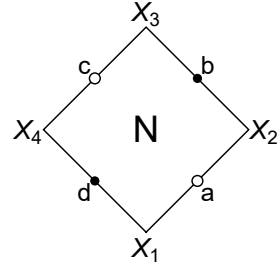}
\caption{\label{fig:appa} An  elementary patch on the worldsheet. $X_i$ are points in spacetime, $N$ is the normal vector of the patch. The variables $a,b,c,d$ are those that appear in eqn. (\ref{eq:deqn}).
}
\end{center}
\end{figure}

Here we list some formulas that allow one to convert the discrete field  into spacetime coordinates. FIG. \ref{fig:appa} shows a single elementary patch on the worldsheet. Its four vertices are denoted by $X_i \in \RR^{2,2}$. The patch is characterized by a constant normal vector $N \in \RR^{2,2}$. The patch is surrounded by four kinks whose worldlines are null. For instance, $ (X_1 - X_2)^2 = 0$. These difference vectors can then be written as products of helicity spinors as in \cite{Vegh:2016hwq}: $X_1 - X_2 = \lambda \tilde \lambda$. Both spinors can be chosen to have two real components (this is possible because the ambient spacetime is $\RR^{2,2}$). Thus, each spinor is characterized by a magnitude and a phase. The tangents of the phases are denoted $a,b,c,d$ and they sit on the kink lines in  FIG. \ref{fig:appa}. These are the values of the discrete field $a_{ij}$ of eqn. (\ref{eq:deqn}) at four neighboring lattice points. The formula for computing them from the difference vector $X_1 - X_2 \equiv \{ p_{-1}, p_{0}, p_1,  p_2 \}$ is
\be
  \nonumber
  a = {p_{-1}+p_2 \ov p_0 + p_1}
\ee
and similarly for $b,c$, and $d$.

Let us assume that $a_{ij}$ is given for all values of $i$ and $j$. Let us further assume that $X_2 \equiv \{ Y_{-1}, Y_{0}, Y_1,  Y_2 \}$ is given. Then, $N$ can be determined by
\bea
  \label{eq:detn}
   & N = {1\ov a-b} \times \hskip 6cm  &  \\
  \nonumber
  &
  \hskip -0.8cm  \times \left( \begin{array}{l}
     (1+ab) Y_0 + (ab-1) Y_1  - (a+b) Y_2   \\
     -(1+ab) Y_{-1} + (a+b) Y_1  + (ab-1) Y_2   \\
     (ab-1) Y_{-1} + (a+b) Y_0  - (1+ab) Y_2   \\
     -(a+b) Y_{-1} + (ab-1) Y_0  + (1+ab) Y_1
   \end{array} \right)
   &
\eea
Note that $N^2 = 1$ and thus the normal vector has three degrees of freedom. Two of them are fixed by $a$ and $b$, and the third one by the constraint that $X_2$ lies on the patch, i.e. $N \cdot X_2 =0$.

Once $N \equiv \{ n_{-1}, n_{0}, n_1,  n_2 \}$ is determined, the other three $X_i$ vertices of the patch can be determined by inverting (\ref{eq:detn}). For instance,
\bea
  \nonumber
   & X_4 = {1\ov c-d} \times \hskip 6cm  &  \\
  \nonumber
  &
  \hskip -0.8cm  \times \left( \begin{array}{l}
     (1+cd) n_0 + (cd-1) n_1  - (c+d) n_2   \\
     -(1+cd) n_{-1} + (c+d) n_1  + (cd-1) n_2   \\
     (cd-1) n_{-1} + (c+d) n_0  - (1+cd) n_2   \\
     -(c+d) n_{-1} + (cd-1) n_0  + (1+cd) n_1
   \end{array} \right)
   &
\eea

\clearpage

\begin{figure}[h]
\begin{center}
\includegraphics[width=5cm]{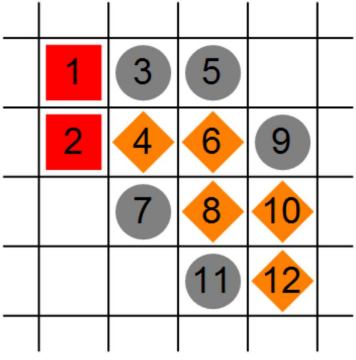}
\caption{\label{fig:appb}
Elementary patches on the worldsheet.
}
\end{center}
\end{figure}

\section*{Appendix B: Initial conditions}

In this section I would like to outline the shooting method I used for finding initial conditions for the calculations. An example plot of the worldsheet is displayed in FIG. \ref{fig:appb}. The horizontal and vertical axes are the lightcone coordinates on the worldsheet.
The black lines are kink worldlines. In between the kink lines we have elementary patches with constant normal vectors. The goal is to find the correct normal vectors of two ``rows'' which allows the simulation to start.
In the figure the first row consists of 2, 7, and 11. The second row is 1, 4, 8, and 12.
This example has only $3+4 = 7$ elementary segments on the string (patch 1 and 12 are both ending on the boundary of AdS$_3$).

The first two patch normals (red squares) are given. Then the numbering shows the order in which we compute the other normal vectors. Gray circles indicate that the normal vector is obtained via a $\delta t$ time-shift on the \poincare patch. If the original normal vector is $N = \{ n_{-1}, n_0, n_1, n_2 \}$, then the time-shifted normal vector is
\be
   \nonumber
 N' =\left( \begin{array}{c}
     n_{-1} + \delta t  \, n_0 - \half \delta t^2 (n_{-1}+n_2)   \\
     n_0 - \delta t (n_{-1}+n_2)  \\
     n_1  \\
     n_2-\delta t \, n_0  +  \half \delta t^2 (n_{-1}+n_2)
   \end{array} \right)
\ee
Using this formula, we compute $N_3$ from $N_2$. The time-shift parameter is a variable at this point and all the normal vectors $N_i$ for $i\ge 3$ will depend on it.

From $N_1, N_2$, and $N_3$, we can compute $N_4$ by the reflection formula
\be
   \nonumber
 N_4 = -N_1 + 2{N_2+N_3 \ov 1+N_2 \cdot N_3}
\ee
The fact that $N_4$ has been computed via the reflection formula is indicated by an orange diamond. Now we compute $N_5$ again with a $\delta t$ time-shift from $N_4$. Then we move on to compute $N_6$ by the reflection formula.

\begin{figure}[h]
\begin{center}
\includegraphics[width=5cm]{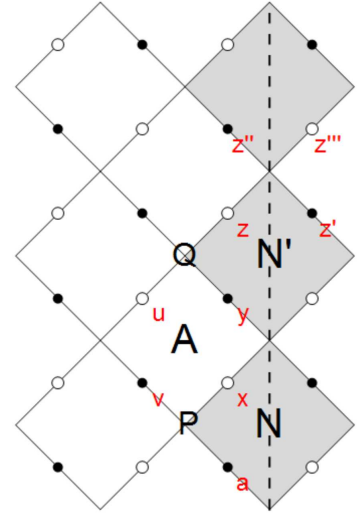}
\caption{\label{fig:appc}
Boundary conditions. Vertical and horizontal directions are $\tau$ and $\sigma$, respectively.  Kink worldlines are shown as solid black lines. Vertical dashed line indicates the location where the worldsheet touches the AdS$_3$ boundary. $P, Q$ are spacetime points, while $A, N$, and $N'$ are normal vectors. Red lowercase letters label the discrete $a_{ij}$ field at different points.
}
\end{center}
\end{figure}

The remaining normal vectors are computed in a similar fashion. Once we reach the required number of segments, we examine the final normal vector -- in this case $N_{12}$ -- and adjust $\delta t$ via a shooting method such that $N_{12}$ is again perpendicular to the boundary. Finally, we rescale the entire string such that the endpoints on the \poincare patch are at $x=\pm 1$, and convert the normal vectors into the $a_{ij}$ variables.

\section*{Appendix C: Boundary conditions}

This Appendix discusses the details of enforcing the correct boundary conditions. This is crucial for calculation. FIG. \ref{fig:appc} shows the worldsheet which ends on the boundary of AdS$_3$ along the vertical dashed line.

Let us first assume that we want fixed Dirichlet boundary conditions for this string endpoint. This means that we want $N'$ simply be a time-shift of $N$ (see Appendix B for the time-shift formula). Furthermore, let us assume that we have already computed three rows of the discrete field (red letters in the figure), and thus we know $a, x, y, u$, and $v$. We cannot compute $z$ using eqn. (\ref{eq:deqn}), since one of the variables is missing (i.e. the mirror image of $y$). Hence its value is determined by the boundary condition. Now $N$ is known, and using the results of Appendix B, we can compute the spacetime point $P = P(N, a, x)$. Then we compute the normal vector $A = A(P, x,v)$, then $Q = Q(A, u,y)$, and finally we have $N' = N'(Q, z, y)$. If we now let this equal to a time-shifted $N$ vector, then we get an equation for $z$.

\begin{figure}[h]
\begin{center}
\includegraphics[width=10cm]{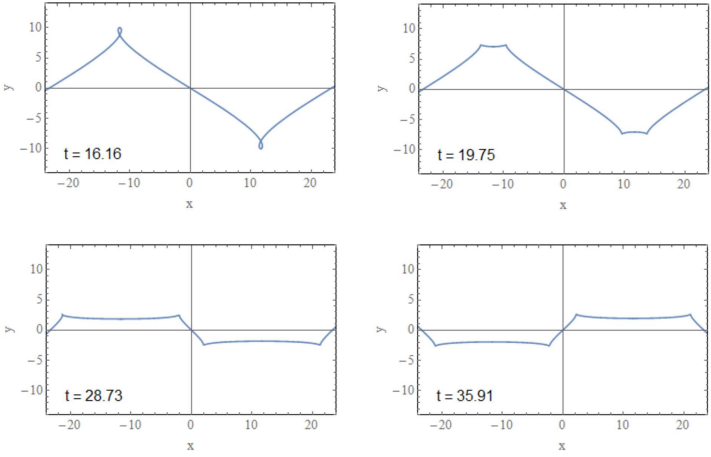}
\caption{\label{fig:string}
Timeslices of an analytical solution in flat space. The first plot has no cusps. Cusps in the other three plots move with the speed of light.
}
\end{center}
\end{figure}

Finally, $z'' = z$ since the corresponding kink lines touch the boundary at the same time (and this is the actual meaning of these variables).

If the endpoint is not fixed (for instance, when we are creating the initial wave on the string), then $N'$ will not be a simple time-shift of $N$. In this case, $z$ has to be set by a shooting method such that $N'$ best approximates the prescribed boundary condition which is a smooth function in our case.

Note that this method of setting the boundary condition requires three rows of the discrete field. Since initial conditions only provide two rows, one has to make sure that these two rows are those of $a$ and $x$ in the figure, because then $y = x$ is easily computed. If we were to start out with the rows of $x$ and $y$ (without knowing the values of the field in the row of $a$), then we would not know how to proceed.


\section*{Appendix D: An analytic solution in flat space}

This Appendix contains an analytical solution in flat space which shares some features with the non-linear AdS$_3$ solutions in the bulk of the paper. The string embedding is explicitly given by
\bea
  \nonumber
  X^t(\tau, \sigma) = &  {1 \over 4 \sqrt{2}} \le( {(8+2a^2)\tau - a^2 \cos 2\sigma \sin 2\tau } \ri) & \\
   \nonumber
  X^x(\tau, \sigma) =  &  {1 \over 4 \sqrt{2}} \le( {(8-2a^2)\sigma + a^2 \sin 2\sigma \cos 2\tau } \ri)  & \\
  \nonumber
  X^y(\tau, \sigma) =     &   2a \sin\sigma \cos\tau  & .
\eea

The solution satisfies the flat space equation of motion and the Virasoro constraints,
\be
  \nonumber
  \p \bar\p \vec X = 0 \, , \quad  (\p \vec X)^2 =(\bar \p \vec X)^2 =  0
\ee

\begin{figure}[h]
\begin{center}
\includegraphics[width=6cm]{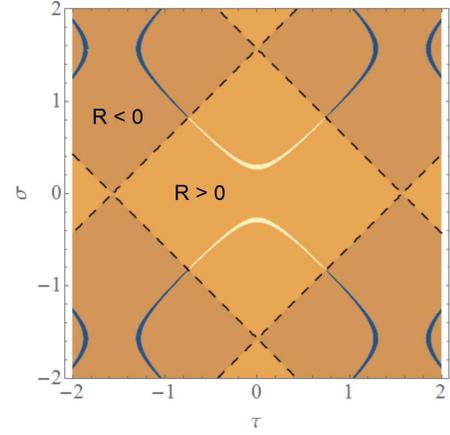}
\caption{\label{fig:flatricci}
Induced scalar curvature of the analytical solution in flat space (with $a=5$).  Dashed lines indicate where the curvature changes sign. Cusps on the string correspond to the blue/white curves in the figure. The worldlines are subluminal (superluminal) in the $R>0$ ($R<0$) regions.
}
\end{center}
\end{figure}

\noindent
where $z=\tau+\sigma$ and $\bar z=\tau-\sigma$.
FIG. \ref{fig:string} shows the string at different moments of time. The first plot contains no cusps which then later appear. The motion is time-periodic, and thus the cusps will finally recombine and disappear.

Note that in the second plot of FIG. \ref{fig:string}  the newly minted cusps (at $x \approx 10$ and  $ x \approx 14$) are on the same side of the string. This is because the string in the first plot has a loop and the associated topological charge has to be preserved under time-evolution.
In FIG. \ref{fig:cusppair}, cusps were on the opposite sides, because the string did not contain such a loop before the moment of pair-creation.

The scalar curvature of the induced metric can be easily computed. It is given by
\be
  \label{eq:flatricci}
  R = {256 a^2 (\cos 2\sigma + \cos 2\tau) \ov (4+a^2 \cos 2\sigma -a^2 \cos 2\tau)^4}
\ee

A density plot of this function is shown in FIG. \ref{fig:flatricci}.
The brighter and darker regions indicate the sign of $R$. These are analogous to the shG and chG regions in the AdS$_3$ case.

As the figure displays, at $\tau=0$ the singular objects are momentarily static. Let us pick one of the singularities which is located at
\be
  \nonumber
  \sigma = \sigma_0 \equiv \half \arccos {a^2-4 \ov a^2}
\ee
Expanding the curvature (\ref{eq:flatricci}) near $\sigma_0$ we get
\be
  \label{eq:riccilimit}
   R \propto (\sigma-\sigma_0)^{-4}
\ee
In order to see what this has to do with sinh-Gordon solitons, let us expand the exact static soliton solution in (\ref{eq:solitt}) near its singularity. We get
\be
  \nonumber
   \alpha \propto \log(\sigma-\sigma_0)
\ee
By applying (\ref{eq:ricci}), the scalar curvature is seen to have the same singular scaling as in (\ref{eq:riccilimit}). This implies that this singular object in flat space is indeed analogous to the well-known sinh-Gordon soliton.

\bibliography{broken}

\begin{thebibliography}{31}
\expandafter\ifx\csname natexlab\endcsname\relax\def\natexlab#1{#1}\fi
\expandafter\ifx\csname bibnamefont\endcsname\relax
  \def\bibnamefont#1{#1}\fi
\expandafter\ifx\csname bibfnamefont\endcsname\relax
  \def\bibfnamefont#1{#1}\fi
\expandafter\ifx\csname citenamefont\endcsname\relax
  \def\citenamefont#1{#1}\fi
\expandafter\ifx\csname url\endcsname\relax
  \def\url#1{\texttt{#1}}\fi
\expandafter\ifx\csname urlprefix\endcsname\relax\def\urlprefix{URL }\fi
\providecommand{\bibinfo}[2]{#2}
\providecommand{\eprint}[2][]{\url{#2}}

\bibitem[{\citenamefont{Maldacena}(1998)}]{Maldacena:1997re}
\bibinfo{author}{\bibfnamefont{J.~M.} \bibnamefont{Maldacena}},
  \bibinfo{journal}{Adv. Theor. Math. Phys.} \textbf{\bibinfo{volume}{2}},
  \bibinfo{pages}{231} (\bibinfo{year}{1998}).

\bibitem[{\citenamefont{Gubser et~al.}(1998)\citenamefont{Gubser, Klebanov, and
  Polyakov}}]{Gubser:1998bc}
\bibinfo{author}{\bibfnamefont{S.~S.} \bibnamefont{Gubser}},
  \bibinfo{author}{\bibfnamefont{I.~R.} \bibnamefont{Klebanov}},
  \bibnamefont{and} \bibinfo{author}{\bibfnamefont{A.~M.}
  \bibnamefont{Polyakov}}, \bibinfo{journal}{Phys. Lett.}
  \textbf{\bibinfo{volume}{B428}}, \bibinfo{pages}{105} (\bibinfo{year}{1998}),
  \eprint{hep-th/9802109}.

\bibitem[{\citenamefont{Witten}(1998)}]{Witten:1998qj}
\bibinfo{author}{\bibfnamefont{E.}~\bibnamefont{Witten}},
  \bibinfo{journal}{Adv. Theor. Math. Phys.} \textbf{\bibinfo{volume}{2}},
  \bibinfo{pages}{253} (\bibinfo{year}{1998}).

\bibitem[{\citenamefont{Ishii and Murata}(2015)}]{Ishii:2015wua}
\bibinfo{author}{\bibfnamefont{T.}~\bibnamefont{Ishii}} \bibnamefont{and}
  \bibinfo{author}{\bibfnamefont{K.}~\bibnamefont{Murata}},
  \bibinfo{journal}{JHEP} \textbf{\bibinfo{volume}{06}}, \bibinfo{pages}{086}
  (\bibinfo{year}{2015}), \eprint{1504.02190}.

\bibitem[{\citenamefont{Pohlmeyer}(1976)}]{Pohlmeyer:1975nb}
\bibinfo{author}{\bibfnamefont{K.}~\bibnamefont{Pohlmeyer}},
  \bibinfo{journal}{Commun. Math. Phys.} \textbf{\bibinfo{volume}{46}},
  \bibinfo{pages}{207} (\bibinfo{year}{1976}).

\bibitem[{\citenamefont{De~Vega and Sanchez}(1993)}]{DeVega:1992xc}
\bibinfo{author}{\bibfnamefont{H.~J.} \bibnamefont{De~Vega}} \bibnamefont{and}
  \bibinfo{author}{\bibfnamefont{N.~G.} \bibnamefont{Sanchez}},
  \bibinfo{journal}{Phys. Rev.} \textbf{\bibinfo{volume}{D47}},
  \bibinfo{pages}{3394} (\bibinfo{year}{1993}).

\bibitem[{\citenamefont{Bena et~al.}(2004)\citenamefont{Bena, Polchinski, and
  Roiban}}]{Bena:2003wd}
\bibinfo{author}{\bibfnamefont{I.}~\bibnamefont{Bena}},
  \bibinfo{author}{\bibfnamefont{J.}~\bibnamefont{Polchinski}},
  \bibnamefont{and} \bibinfo{author}{\bibfnamefont{R.}~\bibnamefont{Roiban}},
  \bibinfo{journal}{Phys. Rev.} \textbf{\bibinfo{volume}{D69}},
  \bibinfo{pages}{046002} (\bibinfo{year}{2004}), \eprint{hep-th/0305116}.

\bibitem[{\citenamefont{Gubser et~al.}(2002)\citenamefont{Gubser, Klebanov, and
  Polyakov}}]{Gubser:2002tv}
\bibinfo{author}{\bibfnamefont{S.~S.} \bibnamefont{Gubser}},
  \bibinfo{author}{\bibfnamefont{I.~R.} \bibnamefont{Klebanov}},
  \bibnamefont{and} \bibinfo{author}{\bibfnamefont{A.~M.}
  \bibnamefont{Polyakov}}, \bibinfo{journal}{Nucl. Phys.}
  \textbf{\bibinfo{volume}{B636}}, \bibinfo{pages}{99} (\bibinfo{year}{2002}),
  \eprint{hep-th/0204051}.

\bibitem[{\citenamefont{Kruczenski}(2005)}]{Kruczenski:2004wg}
\bibinfo{author}{\bibfnamefont{M.}~\bibnamefont{Kruczenski}},
  \bibinfo{journal}{JHEP} \textbf{\bibinfo{volume}{08}}, \bibinfo{pages}{014}
  (\bibinfo{year}{2005}), \eprint{hep-th/0410226}.

\bibitem[{\citenamefont{Jevicki et~al.}(2008)\citenamefont{Jevicki, Jin,
  Kalousios, and Volovich}}]{Jevicki:2007aa}
\bibinfo{author}{\bibfnamefont{A.}~\bibnamefont{Jevicki}},
  \bibinfo{author}{\bibfnamefont{K.}~\bibnamefont{Jin}},
  \bibinfo{author}{\bibfnamefont{C.}~\bibnamefont{Kalousios}},
  \bibnamefont{and} \bibinfo{author}{\bibfnamefont{A.}~\bibnamefont{Volovich}},
  \bibinfo{journal}{JHEP} \textbf{\bibinfo{volume}{0803}}, \bibinfo{pages}{032}
  (\bibinfo{year}{2008}), \eprint{0712.1193}.

\bibitem[{\citenamefont{Jevicki and Jin}(2008)}]{Jevicki:2008mm}
\bibinfo{author}{\bibfnamefont{A.}~\bibnamefont{Jevicki}} \bibnamefont{and}
  \bibinfo{author}{\bibfnamefont{K.}~\bibnamefont{Jin}}, \bibinfo{journal}{Int.
  J. Mod. Phys.} \textbf{\bibinfo{volume}{A23}}, \bibinfo{pages}{2289}
  (\bibinfo{year}{2008}), \eprint{0804.0412}.

\bibitem[{\citenamefont{Dorey and Losi}(2008)}]{Dorey:2008vp}
\bibinfo{author}{\bibfnamefont{N.}~\bibnamefont{Dorey}} \bibnamefont{and}
  \bibinfo{author}{\bibfnamefont{M.}~\bibnamefont{Losi}}
  (\bibinfo{year}{2008}), \eprint{0812.1704}.

\bibitem[{\citenamefont{Jevicki and Jin}(2009)}]{Jevicki:2009uz}
\bibinfo{author}{\bibfnamefont{A.}~\bibnamefont{Jevicki}} \bibnamefont{and}
  \bibinfo{author}{\bibfnamefont{K.}~\bibnamefont{Jin}},
  \bibinfo{journal}{JHEP} \textbf{\bibinfo{volume}{0906}}, \bibinfo{pages}{064}
  (\bibinfo{year}{2009}), \eprint{0903.3389}.

\bibitem[{\citenamefont{Dorey and Losi}(2010{\natexlab{a}})}]{Dorey:2010iy}
\bibinfo{author}{\bibfnamefont{N.}~\bibnamefont{Dorey}} \bibnamefont{and}
  \bibinfo{author}{\bibfnamefont{M.}~\bibnamefont{Losi}}, \bibinfo{journal}{J.
  Phys.} \textbf{\bibinfo{volume}{A43}}, \bibinfo{pages}{285402}
  (\bibinfo{year}{2010}{\natexlab{a}}), \eprint{1001.4750}.

\bibitem[{\citenamefont{Dorey and Losi}(2010{\natexlab{b}})}]{Dorey:2010id}
\bibinfo{author}{\bibfnamefont{N.}~\bibnamefont{Dorey}} \bibnamefont{and}
  \bibinfo{author}{\bibfnamefont{M.}~\bibnamefont{Losi}},
  \bibinfo{journal}{JHEP} \textbf{\bibinfo{volume}{12}}, \bibinfo{pages}{014}
  (\bibinfo{year}{2010}{\natexlab{b}}), \eprint{1008.5096}.

\bibitem[{\citenamefont{Irrgang and Kruczenski}(2013)}]{Irrgang:2012xb}
\bibinfo{author}{\bibfnamefont{A.}~\bibnamefont{Irrgang}} \bibnamefont{and}
  \bibinfo{author}{\bibfnamefont{M.}~\bibnamefont{Kruczenski}},
  \bibinfo{journal}{J.Phys.} \textbf{\bibinfo{volume}{A46}},
  \bibinfo{pages}{075401} (\bibinfo{year}{2013}), \eprint{1210.2298}.

\bibitem[{\citenamefont{Losi}(2010)}]{Losi:2010hr}
\bibinfo{author}{\bibfnamefont{M.}~\bibnamefont{Losi}}, Ph.D. thesis,
  \bibinfo{school}{Cambridge U., DAMTP} (\bibinfo{year}{2010}),
  \eprint{1109.5401}.

\bibitem[{\citenamefont{Vegh}(2018)}]{Vegh:2015ska}
\bibinfo{author}{\bibfnamefont{D.}~\bibnamefont{Vegh}}, \bibinfo{journal}{JHEP}
  \textbf{\bibinfo{volume}{02}}, \bibinfo{pages}{045} (\bibinfo{year}{2018}),
  \eprint{1508.06637}.

\bibitem[{\citenamefont{Callebaut et~al.}(2015)\citenamefont{Callebaut, Gubser,
  Samberg, and Toldo}}]{Callebaut:2015fsa}
\bibinfo{author}{\bibfnamefont{N.}~\bibnamefont{Callebaut}},
  \bibinfo{author}{\bibfnamefont{S.~S.} \bibnamefont{Gubser}},
  \bibinfo{author}{\bibfnamefont{A.}~\bibnamefont{Samberg}}, \bibnamefont{and}
  \bibinfo{author}{\bibfnamefont{C.}~\bibnamefont{Toldo}},
  \bibinfo{journal}{JHEP} \textbf{\bibinfo{volume}{11}}, \bibinfo{pages}{110}
  (\bibinfo{year}{2015}), \eprint{1508.07311}.

\bibitem[{\citenamefont{Vegh}(2015)}]{Vegh:2015yua}
\bibinfo{author}{\bibfnamefont{D.}~\bibnamefont{Vegh}} (\bibinfo{year}{2015}),
  \eprint{1509.05033}.

\bibitem[{\citenamefont{Gubser}(2016)}]{Gubser:2016wno}
\bibinfo{author}{\bibfnamefont{S.~S.} \bibnamefont{Gubser}},
  \bibinfo{journal}{Phys. Rev.} \textbf{\bibinfo{volume}{D94}},
  \bibinfo{pages}{106007} (\bibinfo{year}{2016}), \eprint{1601.08209}.

\bibitem[{\citenamefont{Gubser et~al.}(2016)\citenamefont{Gubser, Parikh, and
  Witaszczyk}}]{Gubser:2016zyw}
\bibinfo{author}{\bibfnamefont{S.~S.} \bibnamefont{Gubser}},
  \bibinfo{author}{\bibfnamefont{S.}~\bibnamefont{Parikh}}, \bibnamefont{and}
  \bibinfo{author}{\bibfnamefont{P.}~\bibnamefont{Witaszczyk}},
  \bibinfo{journal}{JHEP} \textbf{\bibinfo{volume}{07}}, \bibinfo{pages}{122}
  (\bibinfo{year}{2016}), \eprint{1602.00679}.

\bibitem[{\citenamefont{Vegh}(2016{\natexlab{a}})}]{Vegh:2016hwq}
\bibinfo{author}{\bibfnamefont{D.}~\bibnamefont{Vegh}}
  (\bibinfo{year}{2016}{\natexlab{a}}), \eprint{1601.07571}.

\bibitem[{\citenamefont{Vegh}(2016{\natexlab{b}})}]{Vegh:2016fcm}
\bibinfo{author}{\bibfnamefont{D.}~\bibnamefont{Vegh}}
  (\bibinfo{year}{2016}{\natexlab{b}}), \eprint{1603.04504}.

\bibitem[{\citenamefont{Hofman and Maldacena}(2006)}]{Hofman:2006xt}
\bibinfo{author}{\bibfnamefont{D.~M.} \bibnamefont{Hofman}} \bibnamefont{and}
  \bibinfo{author}{\bibfnamefont{J.~M.} \bibnamefont{Maldacena}},
  \bibinfo{journal}{J. Phys.} \textbf{\bibinfo{volume}{A39}},
  \bibinfo{pages}{13095} (\bibinfo{year}{2006}), \eprint{hep-th/0604135}.

\bibitem[{\citenamefont{Bastianello et~al.}(2017)\citenamefont{Bastianello,
  Doyon, Watts, and Yoshimura}}]{Bastianello:2017zby}
\bibinfo{author}{\bibfnamefont{A.}~\bibnamefont{Bastianello}},
  \bibinfo{author}{\bibfnamefont{B.}~\bibnamefont{Doyon}},
  \bibinfo{author}{\bibfnamefont{G.}~\bibnamefont{Watts}}, \bibnamefont{and}
  \bibinfo{author}{\bibfnamefont{T.}~\bibnamefont{Yoshimura}}
  (\bibinfo{year}{2017}), \eprint{1712.05687}.

\bibitem[{\citenamefont{Orfanidis}(1978)}]{PhysRevD.18.3822}
\bibinfo{author}{\bibfnamefont{S.~J.} \bibnamefont{Orfanidis}},
  \bibinfo{journal}{Phys. Rev. D} \textbf{\bibinfo{volume}{18}},
  \bibinfo{pages}{3822} (\bibinfo{year}{1978}).

\bibitem[{\citenamefont{Bobenko and Suris}(2002)}]{BobenkoSuris:2001}
\bibinfo{author}{\bibfnamefont{A.~I.} \bibnamefont{Bobenko}} \bibnamefont{and}
  \bibinfo{author}{\bibfnamefont{Y.~B.} \bibnamefont{Suris}},
  \bibinfo{journal}{Internat. Math. Research Notices}
  \textbf{\bibinfo{volume}{11}}, \bibinfo{pages}{573} (\bibinfo{year}{2002}),
  \eprint{nlin/0110004}.

\bibitem[{\citenamefont{Larsen and Sanchez}(1996)}]{Larsen:1996gn}
\bibinfo{author}{\bibfnamefont{A.~L.} \bibnamefont{Larsen}} \bibnamefont{and}
  \bibinfo{author}{\bibfnamefont{N.~G.} \bibnamefont{Sanchez}},
  \bibinfo{journal}{Phys. Rev.} \textbf{\bibinfo{volume}{D54}},
  \bibinfo{pages}{2801} (\bibinfo{year}{1996}), \eprint{hep-th/9603049}.

\bibitem[{\citenamefont{Neveu and Papanicolaou}(1978)}]{Neveu:1977cr}
\bibinfo{author}{\bibfnamefont{A.}~\bibnamefont{Neveu}} \bibnamefont{and}
  \bibinfo{author}{\bibfnamefont{N.}~\bibnamefont{Papanicolaou}},
  \bibinfo{journal}{Commun. Math. Phys.} \textbf{\bibinfo{volume}{58}},
  \bibinfo{pages}{31} (\bibinfo{year}{1978}).

\bibitem[{\citenamefont{Klotzek and Thies}(2010)}]{Klotzek:2010gp}
\bibinfo{author}{\bibfnamefont{A.}~\bibnamefont{Klotzek}} \bibnamefont{and}
  \bibinfo{author}{\bibfnamefont{M.}~\bibnamefont{Thies}}, \bibinfo{journal}{J.
  Phys.} \textbf{\bibinfo{volume}{A43}}, \bibinfo{pages}{375401}
  (\bibinfo{year}{2010}), \eprint{1006.0324}.

\end{thebibliography}

\end{document}